\documentclass{aa}
\usepackage{psfig}

\def\G{$\Gamma$}

\def\s{$\sigma$}
\def\NH{$N_{\rm H}$ }
\def\be{\begin{equation}}
\def\ee{\end{equation}}
\newcommand{\ltsima} {$\; \buildrel < \over \sim \;$}
\newcommand{\gtsima} {$\; \buildrel > \over \sim \;$}
\newcommand{\lta} {\lower.5ex\hbox{\ltsima}}
\newcommand{\gta} {\lower.5ex\hbox{\gtsima}}
\def\simgt{\lower 2pt \hbox{$\, \buildrel {\scriptstyle >}\over {\scriptstyle
 \sim}\,$}}
\def\approxlt{\mathrel{\hbox{\rlap{\lower.55ex \hbox {$\sim$}}
        \kern-.3em \raise.4ex \hbox{$<$}}}}
\def\approxgt{\mathrel{\hbox{\rlap{\lower.55ex \hbox {$\sim$}}
        \kern-.3em \raise.4ex \hbox{$>$}}}}

\begin{document}

  \thesaurus{03         
             (11.01.2;  
               11.06.2;  
               11.14.1;  
               13.25.2)  
}

\title{ROSAT HRI monitoring of X-ray variability in the NLS1 
galaxy PKS~0558--504}  
\author{M. Gliozzi\inst{1},  Th. Boller\inst{1},  W. Brinkmann\inst{1}
\and W.N. Brandt\inst{2}}
\offprints{mgliozzi@xray.mpe.mpg.de} 
\institute{
Max-Planck-Institut f\"ur extraterrestrische Physik,
         Postfach 1603, D-85740 Garching, Germany
\and Department of Astronomy \& Astrophysics, The Pennsylvania State University, 
525 Davey Lab, University Park, PA 16802, USA}
\date{Received: ; accepted: }
   \maketitle
\markboth{M.~Gliozzi et al.: ROSAT HRI observations of PKS~0558-504}
{M.~Gliozzi et al.: ROSAT HRI observations of two Molonglo quasars}

\begin{abstract}
We present the results of ROSAT High Resolution Imager (HRI) observations
and the survey data of the radio-loud Narrow-Line Seyfert 1 galaxy (NLS1)
PKS~0558--504.
We find strong and persistent X-ray variability on both
short and medium time-scales. The most extreme amplitude variations require
a radiative efficiency exceeding the theoretical maximum for a Schwarzschild 
black hole, suggesting the presence of
a rotating black hole or the influence of relativistic beaming effects.
The spatial analysis rules out the
possibility that the high luminosity and the strong variability are
related to a nearby source.

\keywords{Galaxies: active -- 
Galaxies: fundamental parameters  
-- Galaxies: ISM -- Galaxies: nuclei -- X-rays: galaxies }
   \end{abstract}
%
\section{Introduction}  
Narrow-line Seyfert 1 galaxies are identified by their optical emission 
line properties: the ratio [O III]/H$\beta$ is less than 3 and FWHM H$\beta$ is
less than 2000${~\rm km~s^{-1}}$ (Osterbrock \& Pogge 1985, Goodrich 1989). 
Their optical spectra are also characterized by the presence of strong permitted  
Fe II, Ca II, O I $\lambda$ 8446 lines (Persson 1988). NLS1 exhibit
characteristic features at other wavelengths as well: they are seldom 
radio loud (Ulvestad et al. 1995, Siebert et al. 1999, Grupe et al. 1999,
2000) and
they are usually strong infrared emitters (Moran et al. 1996). In X-rays
NLS1  have been generally found to have extreme spectral and variability
properties that might be related to an extreme value of a fundamental
physical parameter, originating from the vicinity of a supermassive black 
hole (e.g. Brandt \& Boller 1998).

\begin{table} 
\caption{Long term X-ray variability}
\begin{center}
\begin{tabular}{llll}
\hline
\noalign{\smallskip}
Satellite & {\rm \G}& ${\rm L}_{\rm 0.2-2.4~keV}$ & Reference\\
\noalign{\smallskip}
       &      &   $({\rm erg~s^{-1}})$ & \\
\noalign{\smallskip}       
\hline
\hline
\noalign{\smallskip}
\noalign{\smallskip}
EINSTEIN & $2.21^\dag$ & $3.2 \times10^{45}$ & Elvis et al. (1992)\\
\noalign{\smallskip}
\hline
\noalign{\smallskip}
EXOSAT & 2.21& $2.8 \times10^{45}$ & Lawson et al. (1992)\\
\noalign{\smallskip}
\hline
\noalign{\smallskip}
      &  $2.24^{+0.08}_{-0.08}$  & $1.5\times10^{45}$ \\
GINGA$^\ddag$ & &  & Remillard et al. (1991)\\
      &   $1.92^{+0.12}_{-0.12}$ & $2.0\times10^{45}$ \\
\noalign{\smallskip}
\hline
\noalign{\smallskip} ROSAT & $3.1^{+0.05}_{-0.06}$& $5.4\times10^{45}$& Brinkmann et al. (1997)\\
 \noalign{\smallskip}
\hline
\noalign{\smallskip}
ASCA & $2.25^{+0.03}_{-0.03}$& $3.1 \times10^{45}$ &Leighly (1999b)\\ 
\noalign{\smallskip}
\hline 
\end{tabular}
\end{center}
\dag~ Photon index taken from the EXOSAT observations.\\
\ddag~ The values refer to quiescence and peak, respectively.
\end{table}

PKS~0558--504 ($z=0.137, m_{\rm B}=14.97$) is one of the few radio-loud 
NLS1 galaxies ($R_{\rm L}=f_{\rm 5 GHz}/f_{\rm B}\simeq 27$, Siebert et al.
1999). It was optically identified on the basis of X-ray positions
from the High Energy Astronomy Observatory (HEAO-1, Remillard et al. 1986).
A Ginga observation (Remillard et al. 1991) showed an increase of the
X-ray flux by 67\% in 3 minutes, implying that the apparent
luminosity must be enhanced by relativistic beaming.
Further X-ray observations with different satellites have confirmed the
steep X-ray spectrum and high luminosity
of this source, but no more relativistic flares have been presented in
the literature. It is important to search for such flares with an X-ray
imaging detector to definitively rule out the possibility that the Ginga 
data suffered from source confusion.
Tab. 1 summarizes the luminosities, observed by previous
X-ray instruments, converted to the ROSAT
soft X-ray band. The conversion to luminosities in the
0.2--2.4 keV energy band was performed
using PIMMS, assuming Galactic absorption (\NH = $4.39\times10^{20}{~\rm cm^
{-2}}$, Dickey \& Lockman 1990) and a power law
spectral model with photon index {\rm\G} ranging between 2.1 (Remillard et al.
1991) and 3.1 (Brinkmann et al. 1997).
However, possible deviations from a single power law 
or long term spectral changes can lead to systematic uncertainties.
The measured soft X-ray spectrum is rather steep and the medium
energy power laws are considerably flatter, but the sparse data do not
allow determination of whether the source shows spectral steepening 
towards lower energies or whether long term spectral changes occur
during intensity variations.
The luminosities were calculated
by assuming a Friedman cosmology with $H_0=70 {~\rm km ~s^{-1}~Mpc^{-1}}$, 
$q_0=0.5$ and isotropic emission. 

In this paper we report the results of two ROSAT HRI observation campaigns 
taken five months apart (in November 1997 and April 1998) and the survey
PSPC data (September 1990), with the purpose to 
check whether the strong X-ray variability is persistent and whether a 
nearby source contributes to
the X-ray flux. In section 2 we present the observations and the spatial 
analysis. Section 3 deals with the variability of 
PKS~0558--504. Section 4 contains the main conclusions.
\begin{figure}
\psfig{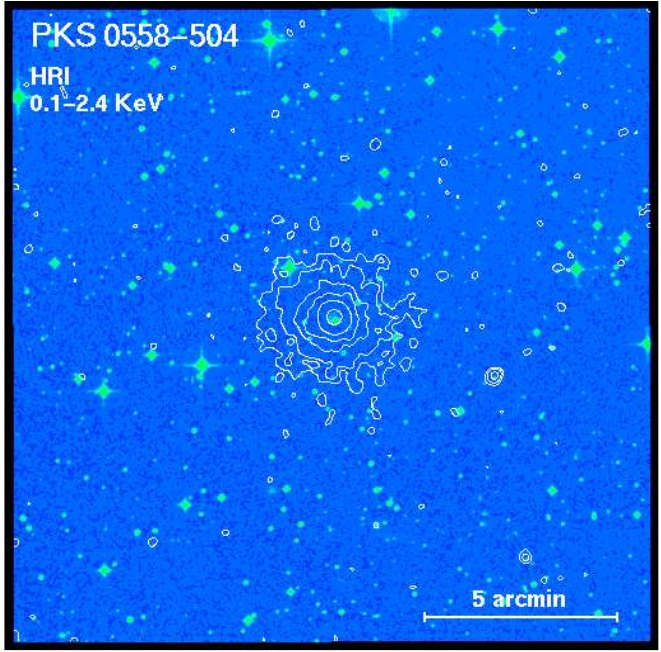}
\caption {Contour plot of the X-ray emission of PKS~0558--504 overlaid
on the optical image of a $16\farcm67 \times 16\farcm67$ region.
The contours correspond to 3, 6, 15, 25, 50 and 3000 \s~ above background.}
\end{figure}

\section{Observations and spatial analysis}

PKS~0558--504 was observed with the ROSAT HRI on November 18 1997, with
an effective exposure of 2.14 ksec, and eleven times between April 19--25, 1998,
with exposures ranging between 860 sec to 4.4 ksec. All individual
observations of April 1998 have been
merged with a final total exposure of 21.52 ksec (see Table 2).
The data analysis was performed using standard routines within the EXSAS
environment (Zimmermann et al. 1994). 
The count rates (vignetting and dead time corrected) quoted in Table 2, 
as well as the light curves, were
obtained by extracting photons from a circle with 150\arcsec~radius around 
the source center and subtracting a background from a source-free region.
In order to reduce the uncertainties from an extrapolation of a
steep power law spectrum to low energies (0.1--0.2 keV) where the Galactic
absorption is important,
we base our discussion on the luminosities in the 0.2--2.4 keV
band.
 As a result the fluxes and luminosities quoted in Table 2 represent 
lower limits only. For completeness, we also mention in the text the values
obtained for the 0.1--2.4 keV energy band, which are typically a factor
of two higher. The conversion factor between HRI
count rates and luminosities, $A_1=3.6\times10^{45}{~\rm erg~count^{-1}}$
($A_2=7.4\times10^{45}{~\rm erg~count^{-1}}$ for 0.1--2.4 keV), was calculated
by assuming a power law spectral model with the best fit parameters 
($\Gamma=2.99$, \NH = $4.53\times 10^{20}{\rm~cm^{-2}}$) of the
PSPC spectrum.
\begin{table} 
\caption{ROSAT HRI observations and results.}
\begin{center}
\begin{tabular}{ccccc}
\hline
\noalign{\smallskip}
Obs. date & Exp. & cts/s & $f_{\rm 0.2-2.4~keV}$ & $L_ {\rm 0.2-2.4~keV}$ \\
~         & [s] &       &
[$\rm erg~cm^{-2} s^{-1}$]& [$\rm erg~s^{-1}$] \\
\noalign{\smallskip}
\hline
\hline
\noalign{\smallskip}
\noalign{\smallskip}
11/18/97 & 2145 & $0.66\pm0.08$ & 
$0.5\times 10^{-10}$ & $2.4 \times 10^{45}$\\
\noalign{\smallskip}
\hline
\noalign{\smallskip}
total 4/98 & 21516 & $1.60\pm0.39$ & 
$1.2\times 10^{-10}$ & $5.7 \times 10^{45}$\\
\noalign{\smallskip}
\hline
\end{tabular}
\end{center}
\end{table}

A peculiar property 
displayed by PKS~0558--504 is the unusually high ratio of X-ray to radio 
luminosity (Brinkmann et al. 1997), that might imply a contribution to 
the X-ray flux from a nearby source.
To perform a spatial analysis we used all the April 1998 observations. 
A contour plot of the X-ray emission overlaid onto the optical image is shown
in Fig. 1. The photons were binned in 2\arcsec$\times$2\arcsec~pixels and 
smoothed with a Gaussian with $\sigma = 6$\arcsec. The surface brightness 
profile is well fitted by the original HRI PSF-model convolved with an
additional Gaussian to allow for the known smearing of the PSF by residual 
wobble motion, which can vary between different observations (Morse 1994).
The equatorial coordinates of the centroid in the HRI image,
computed from a Gaussian fit to the spatial distribution, are
RA(2000)=$5^h59^m47\fs6$, DEC(2000)=$-50^{\rm o}26\arcmin48\arcsec$, in
good agreement with the optical position, taking into account that the 
internal HRI position error is of the order of 5\arcsec.
The only other X-ray source visible in the field of view is
in the  south-west
of PKS~0558--504 at the position RA(2000)=$5^h59^m21\fs2$,
DEC(2000)=$-50^{\rm o}28\arcmin23\arcsec$, with a mean count rate of
0.0021 ${\rm counts~s^{-1}}$, which corresponds to a flux of
$1.4\times10^{-13}{~\rm erg~cm^{-2}~s^{-1}}$ (assuming Galactic absorption
and a power law spectral model with ${\rm\Gamma=2}$). This object is classified as 
`Stellar' with a magnitude of $m_{\rm B}=19.01$ in the digitized COSMOS UKST
southern sky survey. According to Maccacaro et al. (1988) its X-ray to
optical flux ratio suggests that it is an AGN.
This only additional X-ray source in the HRI field of view 
is nearly a factor 1000 fainter than PKS~0558--504, 
and no other strong X-ray source is found in the ROSAT survey within a 
radius of 2$^{\rm o}$. Therefore contributions from a previously unknown
nearby source
to the high luminosity and the strong variability can be ruled out.

\section{X-ray variability}

\begin{figure}
\psfig{figure=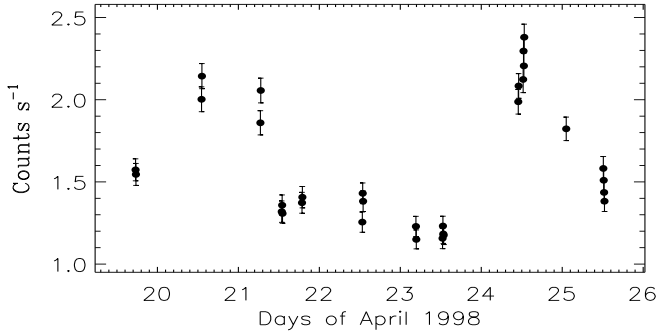,height=4.5cm,width=8.7cm,%
bbllx=60pt,bblly=120pt,bburx=465pt,bbury=373pt,clip=}
\caption{ROSAT HRI light curve for PKS~0558--504 during April 1998 with  
time binning of 400 s.}
\end{figure}

On the basis of the luminosities quoted in Tab. 1, no
long-term X-ray variability by more than a factor of 4 has been seen from 
PKS~0558--504. Recent SAX and RXTE observations seem to confirm this picture
(A. Comastri and K. Leighly, private communication).
However, spectral variations and 
the extrapolation of steep power law spectrum to low energies
can lead to 
uncertain luminosities, as pointed out by Brandt et al. (1999).

More reliable results can be obtained by comparing data  from 
the same instrument. For instance, by comparing the mean count rates of
the two HRI observations taken five months apart, we find an increase
of the count rates
by a factor 2.4, corresponding to a luminosity variation
of ${\rm\Delta}L_{\rm 0.2-2.4~keV}\simeq 3.4\times 10^{45}~{\rm erg~s^{-1}}$
(${\rm\Delta}L_{\rm 0.1-2.4~keV}\simeq 7\times 10^{45}~{\rm erg~s^{-1}}$).

In Fig. 2 we show the total light curve for PKS~0558--504 
during April 1998. The data points are binned into bins of 400 s, in order
to avoid spurious count rate variations due to the ROSAT wobble.
To characterize 
quantitatively the variability in the light curve, we calculated the 
excess variance (Nandra et al. 1997), 
$\sigma^2_{\rm rms}=(5.43\pm2.60)\cdot 10^{-2}$.
At first sight the most extreme count rate variation seems to occur during
April 23 and 24, with ${\rm \Delta cts/\Delta}t= (1.34\pm0.09)$
$\times10^{-5}{\rm counts~s^{-2}}$, corresponding to  
${\rm\Delta}L_{\rm 0.2-2.4~keV}/{\rm\Delta}t=(4.8\pm0.3)
\times 10^{40}~{\rm erg~s^{-2}}$
(${\rm\Delta}L_{\rm 0.1-2.4~keV}/{\rm\Delta}t=(9.9\pm0.7)
\times 10^{40}~{\rm erg~s^{-2}}$),
calculated by performing a linear least square fit to that part 
of the light curve.
However, if we consider the steep increase of the count rate on April 24 only, 
we obtain an even more extreme value of 
${\rm\Delta cts/\Delta}t= (4.13\pm1.34)\times10^{-5}{\rm counts~s^{-2}}$,
leading to ${\rm\Delta}L_{\rm 0.2-2.4~keV}/{\rm\Delta}t=
(1.7\pm0.5)\times 10^{41}~{\rm erg~s^{-2}}$ 
(${\rm\Delta}L_{\rm 0.1-2.4~keV}/{\rm\Delta}t=(3.5\pm1.1)
\times 10^{41}~{\rm erg~s^{-2}}$).
This value can be used to estimate the lower limit of the radiative
efficiency: $\eta > 4.8\times 10^{-43}{\rm\Delta}L/{\rm\Delta}t$ (Fabian 1979).
Straightforward application of the limit gives $\eta > 0.08\pm0.02$
($\eta > 0.17\pm0.05$ for the  0.1--2.4 keV energy band), which exceeds
the theoretical maximum for accretion onto a Schwarzschild
black hole, but not onto a maximally rotating Kerr black hole.
The extremely high efficiency, $\eta\simgt 2$, derived from the Ginga 
2--10 keV luminosities (Remillard et al. 1991) however, strongly indicates
that some approximations used in the calculation of the efficiency limit
must be relaxed, allowing uniform radiation release
and  relativistic effects in the vicinity of the black hole
(e.g.  Brandt et al. 1999).

\begin{figure}
\psfig{figure=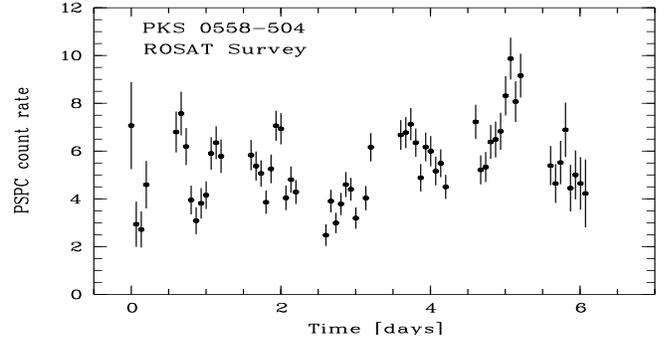,height=4.5cm,width=8.7cm,angle=-90,%
bbllx=45pt,bblly=76pt,bburx=495pt,bbury=660pt,clip=}
\caption{ROSAT All Sky Survey light curve for PKS~0558--504.
\label{figure:lcrass}}
\end{figure}

\begin{figure}
\psfig{figure=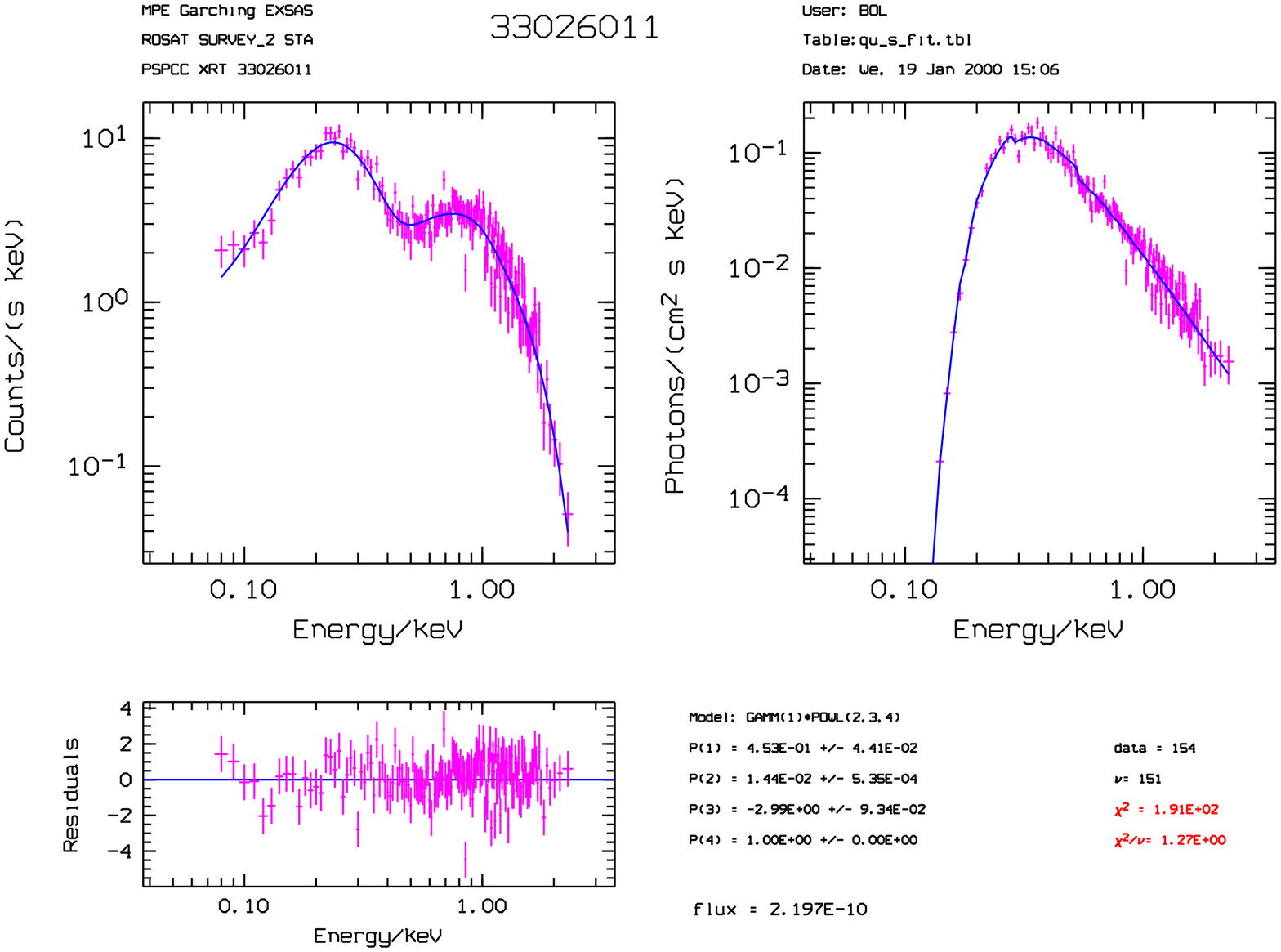,height=4.5cm,width=8.7cm,angle=-90,%
bbllx=408pt,bblly=80pt,bburx=559pt,bbury=394pt,clip=}
\caption{The residuals for power law plus neutral absorption
spectral fits for the survey data of PKS~0558--504.
\label{figure:fitspec}}
\end{figure}

PKS~0558--504 was observed in the ROSAT All-Sky Survey between 1990 
September 8 (10:58:02 UT) and 1990 September 14 (22:14:44 UT), with an 
effective exposure of 1036 s and an average count rate of $\sim 5.5
{\rm~counts~s^{-1}}$. 
The light curve, shown in Fig. \ref{figure:lcrass}
($\sigma^2_{\rm rms}=(6.32\pm0.59)\cdot 10^{-2})$, exhibits variability by more 
than a factor of three with a maximum 
${\rm\Delta cts/\rm\Delta}t= (14.3\pm2.4)$
$\times10^{-5}{\rm counts~s^{-2}}$. 
From the PSPC count rates and  the spectral parameters given below
we obtain
$\Delta L_{\rm0.2-2.4~keV}/\Delta t=(1.8\pm0.3)\times 10^{41}~{\rm erg~s^{-2}}$ 
and
$\Delta L_{\rm0.1-2.4~keV}/\Delta t=(3.6\pm0.6)\times 10^{41}~{\rm erg~s^{-2}}$,
which are very similar to the
values of the most extreme HRI event, but on a different time scale
(the rest frame interval is $\Delta t\sim 7$h for the survey data and
$\Delta t\sim 1.5$h for the HRI flare). As a consequence the values
derived for the radiative efficiency are similar: $\eta > 0.09\pm0.01$
for the 0.2--2.4 keV band 
and $\eta > 0.17\pm0.03$ for the 0.1--2.4 keV band. 

The spectrum can be fit with a single power law with photon index
$\Gamma = 2.99\pm0.09$
and free absorption of $N_H = (4.53\pm0.44)\times10^{20}$cm$^{-2}$,
in excellent agreement with the Galactic value. 
The residuals of this fit, given in Fig. \ref{figure:fitspec}, do not give
strong evidence for deviations from a simple power law although there
are indications at $ E \ge 1~$keV for some spectral changes.
With these parameters the resulting unabsorbed fluxes during the survey
observations are $1.2\times 10^{-10}{~\rm erg~cm^{-2}s^{-1}}$ and
$2.5\times 10^{-10}{~\rm erg~cm^{-2}s^{-1}}$, for the 0.2--2.4 keV and 
0.1--2.4 keV energy bands, respectively, corresponding 
to $L_{0.2-2.4~{\rm keV}} =5.9\times10^{45}{~\rm erg~s^{-1}}$ and
$L_{0.1-2.4~{\rm keV}} =1.2\times10^{46}{~\rm erg~s^{-1}}$.

For several AGN, light curves with large amplitude flares have been
interpreted as indication for non-linear processes (Green 1993, 
Boller et al. 1997, Leighly \& O'Brien 1997). 
Given that the April 1998 light curve of PKS~0558--504 presents
at least two large flares, we searched for non-Gaussianity (and possibly
for non-linearity; see Leighly 1999a for a detailed discussion), by adopting
the Green (1993) method: a time series is non-Gaussian if the ratio of 
its standard deviation to its mean is larger
than unity. Using the data points in Fig. 2 we find that the unweighted
mean count rate is 1.61 ${\rm counts~s^{-1}}$ and the standard deviation
0.39 ${\rm counts~s^{-1}}$ ($\sigma/\bar x=0.24$). As a result
we do not find evidence for non-Gaussian variability. However this method
assumes that the sample mean and standard deviation used are accurate
representations of the true mean and standard deviation, and this might not
be true in our case, due to the limited number of observation intervals.
The same conclusions were reached by 
Leighly (1999a) from ASCA data, using a different method based on the 
skewness of the flux distribution.

\section{Conclusions}
We have presented ROSAT HRI observations of the radio-loud NLS1 galaxy
PKS~0558--504. The main results can be summarized as follows:

From the spatial analysis, no other strong X-ray sources have
been detected in the neighborhood of PKS~0558--504, therefore external
contributions
to the high luminosity and to the strong variability from a
nearby source are ruled out. 
By comparing the X-ray observations throughout the last decade, it 
is evident that the strong X-ray variability of PKS~0558--504 occurs 
persistently.
During the ROSAT HRI observations, PKS~0558--504 shows strong variability, 
both on medium (months) and short (days, hours) time scales. The most extreme
variation implies a radiative efficiency
larger than the theoretical maximum for accretion onto a 
Schwarzschild black hole, and our findings generally support those of Remillard
et al. (1991) where a relativistic $(\eta\simgt 2)$ flare was discovered.
As PKS~0558--504 is a radio-loud object, beamed emission from a jet
could be the cause for the brightness and variability in X-rays.
However, it is worth noting that the radio-quiet NLS1 PHL~1092 has also shown a
relativistic flare $(\eta\simgt 0.6)$, and radio-quiet NLS1 more generally
show enhanced X-ray variability.
The soft X-ray spectrum is rather steep with a power law
 photon index of $\Gamma \sim 3.0$ and 
shows no strong indications for spectral breaks.
The obtained medium energy power laws are considerably flatter 
($\Gamma \sim 2.2$) but the
sparse data and the limited energy bands of the different instruments do 
not allow determination of whether the source shows a spectral steepening towards
lower energies or whether long term spectral changes occur during intensity
variations.  
An answer to these vital questions can only be given by the current broad
band X-ray missions like SAX, XMM-Newton or Chandra.

\begin{acknowledgements}
The ROSAT project is supported by the Bundesministerium f\"ur
Bildung, Wissenschaft, Forschung und Technologie (BMBF) and
the Max-Planck-Gesellschaft.  
MG  acknowledges support from
the European Commission under contract number ERBFMRX-CT98-0195 
(TMR network ``Accretion onto black holes, compact stars and protostars").
WNB acknowledges support from NASA LTSA grant NAG5-8107.
\end{acknowledgements}


\begin{thebibliography}{}

\bibitem{} Boller Th., Brandt W.N., Fabian A.C., Fink H.H., 1997, MNRAS 289, 393

\bibitem{} Brandt W.N., Boller Th., 1998, Astron. Nachr. 319, 7

\bibitem{} Brandt W.N., Boller Th., Fabian A.C., Ruszwoski M., 1999, MNRAS 303,
L53

\bibitem{} Brinkmann W., Yuan W., Siebert J., 1997, A\&A 319, 413

\bibitem{} Dickey J.M., Lockman F.J., 1990, ARA\&A 28, 215

\bibitem{} Elvis M., Plummer D., Schachter J., Fabbiano G., 1992, ApJS 80, 257

\bibitem{} Fabian A.C., 1979, Proc. R. Soc. London, Ser. A 366, 449

\bibitem{} Goodrich R.W., 1989, ApJ 342, 224

\bibitem{} Green A.R., 1993, PhD thesis, Univ. of Southampton

\bibitem{} Grupe D., Beuermann K., Mannhein K., Thomas H.-C., 1999, A\&A
350, 805

\bibitem{} Grupe D., Leighly K. M., Thomas H.-C., Laurent-Muehleisen S. A.,
2000 A\&A in press (astro-ph/0001412)

\bibitem{} Lawson A.J., Turner M.J.L., 1997, MNRAS 288, 920

\bibitem{} Leighly K.M., O'Brien P.T., 1997, ApJ 481, L15

\bibitem{} Leighly K.M., 1999a, ApJS 125, 297

\bibitem{} Leighly K.M., 1999b, ApJS 125, 317

\bibitem{} Maccacaro T., Gioia I.M., Wolter A., Zamorani G., Stocke J.T.,
1988, ApJ 326, 680 

\bibitem{} Moran E.C., Halpern J.P., Helfand D.J., 1996, ApJS 106, 341

\bibitem{} Morse J.A., 1994, PASP 106, 675

\bibitem{} Nandra K., George I.M., Mushotzky R.F., Turner T.J., Yaqoob T.,
1997, ApJ 476, 70

\bibitem{} Osterbrock D.E., Pogge R.W., 1985, ApJ 297, 166

\bibitem{} Persson S.E., 1988, ApJ 330, 751

\bibitem{} Remillard R.A., Bradt H.V., Buckley D.A.H., Roberts W.,
  Schwartz D.A., et al., 1986, ApJ 301, 742

\bibitem{} Remillard R.A., Grossan B., Bradt H.V., Ohashi T., Hayashida K.,
et al., 1991, Nat 350, 589


\bibitem{} Siebert J., Leighly K.M., Laurent-Muehleisen S.A.
Brinkmann W., Boller Th., Matsuoka M., 1999, A\&A 348, 678

\bibitem{} Ulvestad J.S., Antonucci R.R.J. Goodrich R.W., 1995 AJ 109, 81

\bibitem{} Zimmermann H.U., Becker W., Belloni T., et al., 1994, EXSAS User's
Guide, MPE Report 257

\end{thebibliography}
\end{document}